\pdfoutput=1
\documentclass[twocolumn]{aastex61}

\usepackage{savesym}
\savesymbol{tablenum}
\usepackage[separate-uncertainty=true,multi-part-units=single]{siunitx}
\restoresymbol{SIX}{tablenum}
\DeclareSIUnit\year{yr}

\shorttitle{UCLCHEM: A Gas-Grain Chemical Code}
\shortauthors{Holdship et al.}

\begin{document}


\title{UCLCHEM\footnote{\url{https://uclchem.github.io}}: A Gas-Grain Chemical Code for Clouds, Cores and C-Shocks}


\author[0000-0003-4025-1552]{J. Holdship}
\affil{Department of Physics and Astronomy, University College London,    Gower Street, WC1E 6BT}
\email{jrh@star.ucl.ac.uk}

\author{S. Viti}
\affil{Department of Physics and Astronomy, University College London,    Gower Street, WC1E 6BT}

\author{I. Jim\'enez-Serra}
\affil{School of Physics and Astronomy, Queen Mary University of London, 327 Mile End Road, London, E1 4NS}

\author{A. Makrymallis}
\affil{Department of Physics and Astronomy, University College London,    Gower Street, WC1E 6BT}

\author{F. Priestley}
\affil{Department of Physics and Astronomy, University College London,    Gower Street, WC1E 6BT}


%

\begin{abstract}
We present a publicly available, open source version of the time dependent gas-grain chemical code UCLCHEM. UCLCHEM propagates the abundances of chemical species through a large network of chemical reactions in a variety of physical conditions. The model is described in detail along with its applications. As an example of possible uses, UCLCHEM is used to explore the effect of protostellar collapse on commonly observed molecules and to study the behaviour of molecules in C-type shocks. We find the collapse of a simple Bonnor-Ebert sphere successfully reproduces most of the behaviour of CO,CS and NH$_3$ from cores observed by \citet{Tafalla2004} but cannot predict the behaviour of N$_2$H$^+$. In the C-shock application, we find that molecules can be categorized so that they can be useful observational tracers of shocks and their physical properties. Whilst many molecules are enhanced in shocked gas, we identify two groups of molecules in particular. A small number of molecules are enhanced by the sputtering of the ices as the shock propagates and then remain high in abundance throughout the shock. A second, larger set are also enhanced by sputtering but then are destroyed as the gas temperature rises. Through these applications the general applicability of UCLCHEM is demonstrated.
\end{abstract}



\section{Introduction}
Chemistry is ubiquitous in astrophysical environments. Molecular clouds, the cold cores in which stars form, and the warm gas surrounding protostars all exhibit chemistry of varying degrees of complexity and with different dominant chemical pathways. Understanding this chemistry is vital to the study of our own origins as well as understanding the physical structure and processes involved in star formation.\par
Beyond this, chemistry is a useful tool for understanding the physical conditions of the region being studied. This requires well constrained chemical networks and accurate physical models so that uncertainties in the predictions of the model are much smaller than the uncertainties in the measured abundances of molecules. With the current state of the art models, networks are generally capable of putting broad constraints on the physical conditions such as maximum temperatures or minimum densities and this can be of use in poorly understood regions.\par
Chemical modelling is typically performed by the use of rate equations. The rates of a network of chemical reactions are calculated and used to determine the rate of change of a set of chemical species. This coupled set of ordinary differential equations (ODEs) is integrated to give the abundance of each species at any given time. These models typically centre around gas-phase reactions provided by databases such as KIDA \citep{Wakelam2012} and UMIST \citep{McElroy2013}. Each of these databases provides a chemical code, respectively Nahoon and RATE13, which solve these networks for simple gas conditions and include other processes such as H$_2$ formation on the dust grains and the self-shielding of CO and H$_2$ from UV radiation.\par
This, of course, does not account for all astrochemical processes and many groups use more or less complicated chemical models for different purposes. In radiation-hydrodynamic models, it is common to reduce the network to gas-phased H,C and O based species to reduce chemical integration time whilst reproducing the abundances of major coolants such as CO given by more detailed chemical models. On the other hand, the modelling of dense prestellar cores or the formation of complex organic molecules requires much larger chemical networks and the addition of processes involving dust grains.\par
For example, Astrochem \citep{Maret2015} includes freeze out of species onto dust grains and the non-thermal desorption of species from those grains due to UV radiation and cosmic rays making it suitable for modelling a wider range of regions than a simple gas-phase model. By further including thermal desorption, star-forming regions with high gas/dust temperature conditions can be modelled. As the temperature of the core rises, the material on the grains sublimates and proper treatment of this sublimation is required. Astrochem and UCLCHEM implement these processes, with desorption from the grains occurring according to the binding energy of each species onto the grain.\par
However, \citet{Collings2004} demonstrated through temperature programmed desorption experiments that multiple desorption events can occur for each species frozen on the dust grains at different temperatures dependent both on the species and on the rate of change of the temperature. UCLCHEM additionally includes these desorption events for each species on the grain surface (See Section~\ref{sec:thermdesorb}). The modelling of many protostellar environments benefit from the addition of grain chemistry and proper temperature dependent sublimation is required, for instance, for massive hot cores.\par
Another environment where proper grain surface as well as gas-phase chemical treatment has proven useful is in shocked gas. In these systems, sublimation from the grain surface is dominated by sputtering, the collisional removal of surface material. The physical modelling of these shocks is an active area of research \citep[eg.][]{VanLoo2013,Anderl2013} and codes for this modelling are publicly available \citep{Flower2015}. These models have been successful in reproducing observations of shocked regions. For example, L1157-mm is a protostar driving a bipolar outflow \citep{gueth1997} and the bowshocks associated with that outflow are well studied. For one of these bowshocks, L1157-B1, \citet{gusdorf2008} reproduced the SiO emission using an MHD hydrodynamical model of a continuous or C-type shock coupled with a reduced chemical network of 100 species.\par
As noted, reduced chemical networks often produce similar abundances to much more complex networks when simple species such as H$_2$O and CO are considered and this is sufficient for many applications \citep[eg.][]{Schmalzl2014}. Similarly, a simplified physical model may reproduce the shock features that are useful for chemistry and so a full chemical model can be run without fully solving the MHD equations. \citet{jimenez2008} produced a parameterized C-shock that produced a shock structure in good agreement with more complicated MHD models. This was combined with the chemical code UCLCHEM to study the chemistry in shocked environments \citep{viti2011}. Using this parameterization with a large chemical network has been met with some success in the bowshock L1157-B1 \citep{viti2011,holdship2016,Lefloch2016} despite being computationally inexpensive. \par
The simplicity of these models means they could be used to link observed molecular lines with the underlying chemistry to get a sense of the physical properties of the shock. Conversely, emission from a source with well constrained physical properties could be used to improve uncertain parts of the chemical network. The reaction rates of species on dust grains being a prime example. For example, \citet{holdship2016} used observations of the L1157-B1 bowshock to constrain sulfur chemistry on ice grains by identifying likely sulfur carriers. Ideally, chemical models could be ran over large parameter spaces to find the most likely values for uncertain parts of the network when a region has large amounts of observational data available. \par
To this purpose UCLCHEM has been developed over many years with numerous papers discussing major updates and applications \citep[see][]{viti2004,roberts2007}.There have also been benchmarking efforts in the past \citep{viti2001} comparing the code to other similar models. We present here the most up to date \footnote{\url{https://uclchem.github.io}} version, the source code of which is now publicly available for the first time under the MIT license. This paper describes the code in its most recent state and provides example applications of the model. In Section~\ref{sec:uclchem} the code is discussed, in Sections~\ref{sec:collapses} and  \ref{sec:shocks} two examples of the applications of UCLCHEM are given. The paper is summarized in Section~\ref{sec:summary}.
\section{UClCHEM}
\label{sec:uclchem}
UCLCHEM is a time dependent gas-grain chemical model written in modern Fortran. It is primarily a chemical code, focusing on grain chemistry as well as gas-phase reactions. The chemistry includes freeze out, thermal and non-thermal desorption along with a gas-phase and user-provided grain surface reaction network. In addition, hard-coded treatments for H$_2$ formation on the grains and the self-shielding of CO and H$_2$ are contained in the chemistry module. The chemical solver calculates the rates of all the above processes to follow the fractional abundances of molecules for parcels of gas.\par
UCLCHEM makes use of modern Fortran's modules to separate the chemistry and physics. Interchangeable physics modules control the number and physical conditions of gas parcels used in the chemistry to allow different scenarios to be modelled. In particular, the code is packaged with modules for molecular clouds, hot cores, C-shocks and the post-processing of hydrodynamic simulations. In this section, the chemical model is explained along with an overview of the physical models available and the python code produced to create networks automatically.
\subsection{Chemical Model}
\label{sec:chemmod}
At its core, UCLCHEM is a code that sets up and solves the coupled system of ODEs that gives the fractional abundances of all the species in a parcel of gas. The ODEs are created automatically from the network (see Sec.~\ref{sec:makerates}). There is one ODE per species and each is a sum over the rates of every reaction that involves the species. At each time step, the rates of each reaction are recalculated and the third party ODE solver DVODE \citep{Brown1989} integrates to the end of the time step.
Each type of reaction requires a different rate calculation and so UCLCHEM is limited by the kinds of reactions that have been coded for. The subsections below detail each type of reaction and physical process currently included.
\subsubsection{Gas Phase Reactions}
Gas phase reactions make up the largest part of the chemical network. Two body reactions, cosmic ray interactions and interaction with UV photons are all included in the code. The rate of each reaction is in $cm^{3} s^{-1}$. For two body reactions, the rate is calculated through the Arrhenius equation,
\begin{equation}
R_{AB}=\alpha\left(\frac{T}{300 K}\right)^{\beta}\exp\left({\frac{-\gamma}{T}}\right),
\end{equation}
where $\alpha$, $\beta$ and $\gamma$ are experimentally determined rate constants. For cosmic ray protons, cosmic ray induced photons and UV photons the rates are given by,
\begin{eqnarray}
R_{CRProton}=\alpha_{cr} \zeta \\
R_{CRPhoton}=\alpha_{cr}\left(\frac{T}{300 K}\right)^{\beta}\frac{E}{1-\omega}\zeta\\
R_{UV}=\alpha_{rad} F_{UV}\exp(-kA_v)
\end{eqnarray}
where $\omega$ is the dust grain albedo, $\zeta$ is the cosmic ray ionisation rate and F$_{UV}$ is the UV flux. $\alpha_{rad}$ and $\alpha_{cr}$ scale the overall radiation field and cosmic ray ionisation rates for a specific reaction, $\beta$ takes the same meaning as above, $E$ is the efficiency of cosmic ray ionisation events  and $k$ is a factor increasing extinction for UV light. This formulation of the reaction rates is taken from the UMIST database notes and more information can be found in \citet{McElroy2013}. These rates are then used to set up and solve a series of ordinary differential equations of the form,
\begin{equation}
\dot Y_{product}=R_{AB}Y_AY_Bn_H,
\end{equation}
where $Y$ is the abundance of reactants A and B and $\dot Y$ is the rate of change of the product. $n_H$ is the hydrogen number density. The second abundance dependency is dropped for reactions between single species and photons or cosmic rays. An equivalent negative value is added to the change in the reactants.\par
In addition to these reactions, the rate of dissociation of CO and H$_2$ due to UV is reduced by a self-shielding treatment in the code. H$_2$ formation is treated with the hard-coded reaction rate,
\begin{equation}
R_{H2}=10^{-17} \sqrt{T}n_H Y_H,
\end{equation}
where $Y_H$ is the abundance of atomic hydrogen. This rate is taken from \citet{Dejong1977} and performed well in chemical benchmarking tests \citep{Rollig2007}. It should be noted that, despite the self-shielding treatment and inclusion of UV photon reactions from UMIST, UCLCHEM is not a PDR code and should not be used to model regions where the UV field is sufficient to be considered a PDR. To model the chemistry in PDRs, the 3DPDR code described in \citet{Bisbas2012} is available on the UCLCHEM website and a updated version of the 1D UCL\_PDR \citep{Bell2006} will be available soon.
\subsubsection{Freeze out}
All species freeze out at a rate given by,
\begin{equation}
R_{fr}=\alpha_f\sqrt{\frac{T}{m}}a_gfr,
\end{equation}
where  $a_g$ is the grain radius and $fr$ is the proportion of each species that will freeze. $\alpha_f$ is a branching ratio allowing the same species to freeze through different channels into several different species. Therefore, $\alpha_f$ allows the user to determine what proportion of a species will freeze into different products. This is used as a proxy for the relatively uncertain surface chemistry. For example, by freezing a portion of a species as a more hydrogenated species rather than explicitly including a hydrogenation reaction in the surface network. The values for $\alpha_f$ in each freeze out reaction should sum to 1 for any given species. $fr$ allows the user to set the freeze out efficiency; it is left at 1.0 in this work so that non-thermal desorption alone accounts for molecules remaining in the gas-phase. An additional factor, $C_{ion}$ is included for positive ions to reproduce the electrostatic attraction to the grains,
\begin{equation}
C_{ion}=1+\frac{16.47\times10^{-4}}{a_gT},
\end{equation}
with $a_g$ and $T$ being the average grain radius and temperature respectively and the constant value taken from \citet{Rawlings1992}.
\subsubsection{Non-thermal Desorption}
From \citet{roberts2007}, three non-thermal desorption methods have been included in UCLCHEM. Molecules can leave the grain surface due to the energy released in H$_2$ formation, incident cosmic rays, and by UV photons from the interstellar radiation field as well as those produced when cosmic rays ionise the gas. 
\begin{eqnarray}
R_{DesH2}=R_{H2 form}\epsilon_{H2} M_x,\\
R_{DesCR}=F_{cr}A\epsilon_{cr} M_x,\\
R_{DesUV}=F_{uv}A\epsilon_{UV} M_x
\end{eqnarray}
Where $R_{H2 form}$ is the rate of H$_2$ formation, $F_{cr}$ is the flux of iron nuclei cosmic rays and $F_{uv}$ is the flux of ISRF and cosmic ray induced UV. The $\epsilon$ values are efficiencies for each process reflecting the number of molecules released per event. $A$ is the total grain surface area and $M_x$ is the proportion of the mantle that is species $x$. The values of these parameters are given in Table~\ref{tab:parameters} and the assumptions used to obtain them are discussed in \citet{roberts2007}.
\subsubsection{Thermal Desorption}
\label{sec:thermdesorb}
As the temperature of the dust increases, species start to desorb from the ice mantles. Laboratory experiments \citep[eg.][]{Ayotte2001,Burke2010} show that a species does not desorb in a single peak but rather in multiple desorption events. These events correspond to the removal of the pure species from the mantle surface, a strong peak at the temperature corresponding to the binding energy, and the ``volcanic'' desorption and co-desorption of molecules that have diffused into the water ice. \citet{Collings2004} showed that many species of astrochemical relevance could be classified as CO-like, H$_2$O-like or intermediate, depending on what proportion of the species is removed from the mantle in each desorption event.\par
To the authors knowledge, UCLCHEM is the only publicly available chemical model that implements a treatment for this. Molecules are classified according to their similarity to H$_2$O or CO, controlling their desorption behaviour. Further, the user can now set the proportion of each molecule that enters the gas phase in each event, allowing the thermal desorption treatment of each molecule to improve with the laboratory data available. The user only needs to compile a list of species, along with the binding energies and desorption fractions for each grain species, the rest of this process is set up automatically (see Sect.~\ref{sec:makerates}). Whilst any physics module could activate the desorption process at specific temperatures, this process is strongly linked to the cloud model discussed in Sect.~\ref{sec:cloud}.

\subsubsection{Grain Surface Reactions}
Proper treatment of reactions between molecules on the grain surfaces is subject of debate. As a result, UCLCHEM has implemented grain chemistry in three ways. The method used in the basic network supplied with the code, and in the network used for the applications below, is to include hydrogenation at freeze out. For example, 10\% of CO freezing onto the grains is assumed to go on to form CH$_3$OH in the network used here and so it is immediately added to the grain abundance of CH$_3$OH.\par
More complicated treatments are possible without editing the code. By including reactions with modified version of the rate constants in the network, grain surface reactions can be treated in the same way as gas-phase reactions. Further, UCLCHEM calculates the rates of reactions by diffusion across the grain surfaces for reactions with a third reactant labelled "DIFF". In this case, the binding energy of the two actual reactants are used to calculate their diffusion rates and included in a modified rate equation according to the treatment of \citet{Occhiogrosso2014}.
\subsection{Physical Model}
A number of physics modules are included with the code and can be interchanged as required. They are explained briefly below.\par
\subsubsection{Clouds and Collapse Modes}
\label{sec:cloud}
The ``cloud'' module is the main physics module of UCLCHEM. This sets up a line of parcels from the centre to the edge of a cloud of gas and controls the density, temperature and visual extinction of each effectively giving a 1D model of a cloud. The conditions at the outer edge of the cloud are set by an interstellar radiation field and a value of visual extinction at the cloud edge. The distance from the edge of the cloud and the densities of the parcels closer to the edge are used to calculate the visual extinction and hence the radiation field for each interior parcel but parcels are otherwise treated separately.\par
The model works in two phases. In phase 1, this module is most often used to follow the collapse of the cloud from a diffuse medium to the density required for phase 2. Starting from elemental abundances only, this produces a set of gas and grain molecular abundances that is self consistent with the network. These can be used as the starting values for phase 2 rather than assuming a set of initial abundances. \par
In phase 2, the temperature increases as the cloud collapses and so the envelope around a forming protostar can be modelled. As discussed above, sublimation of species from the grain is dependent both on the temperature and rate of change of temperature.  In the cloud model, the temperature of the gas as a function of time and radial distance from the protostar is calculated at each time step, and compared to pre-calculated temperatures to determine when major thermal desorption events should occur. The temperature profiles and desorption peaks are from \citet{viti2004} where the time-dependent temperature profiles and desorption temperatures for a range of high stellar masses were calculated. This has since been extended with radial dependency and lower masses. Alternatively, a module that reads and interpolates the output of hydrodynamical codes can be used to post-process the chemistry of hydrodynamical models.\par
Phase 1 can also be used to study cold gas. Gas in a steady state can be modelled by setting all the required parameters and turning collapse off. Further, a number of different collapse modes have been coded into the model. The standard collapse used to create a cloud for phase 2 is the freefall collapse taken from \citet{Rawlings1992}. Parameterizations for the collapse of a Bonnor-Ebert sphere are also included so that the collapse of cold cores can be studied. These are described in more detail in Priestley et al. (in prep.). Each collapse model calculates the density of a parcel by following the central density of the Bonnor-Ebert sphere with time and then scaling for the radial distance of a parcel from the centre of the core. The chemistry of each parcel is evolved separately so chemical abundances as a function of radius can be studied for different collapse modes and compared to observations of collapsing cores.
\subsubsection{C-shock}
\label{sec:cshock}
The ``C-shock'' model is the parameterized model of continuous or C-type shocks from \citet{jimenez2008} adapted for use in UCLCHEM. The parameterization calculates the evolution of the velocities of the ion and neutral fluids as a function of time and deduces the physical structure of the shock from simple principles and approximations. In addition to the changes in density and temperature, the shock model includes a treatment of sputtering. The saturation time, t$_{sat}$, is defined in Jimenez-Serra et al. (2008). It corresponds to the time at which the abundances of species sputtered from the icy mantles change by less than 10\% in two consecutive time steps in their calculations. This gives a measurement of the time-scales at which most of the material that should be sputtered from dust grains will have been released. As the timescales in which the mantles sputter are short in the typical conditions in molecular clouds \citep{jimenez2008} this is treated by releasing all of the mantle material into the gas phase at t$_{sat}$. The equations used to calculate the saturation time do not depend on molecular mass and so t$_{sat}$ is the same for all species. See \citet{jimenez2008} Appendix B for more details.\par
Whilst sputtering is included, there are two major grain processes that are omitted. Grain-grain collisions in the shock can cause both shattering and vaporisation. Shattering is the breaking of grains in collisions and primarily has the effect of changing the dust grain size distribution. This produces hotter and thinner shocks, particularly for shocks with higher speeds or weak magnetic fields \citep{Anderl2013}. Vaporisation is the desorption of mantle species due to grain heating from collisions. The inclusion of grain-grain processes vastly increases the complexity of the code and would not allow the code to be run so quickly with low computing power. Shattering in particular is likely to greatly affect the abundance of Si-bearing species as the destroyed dust grains would add to the gas phase abundances. It should be noted however, that in models where SiO is included in the mantle, \citet{Anderl2013} find SiO abundances that are consistent with this model. The interested reader is referred to recent work on grain-grain processes in shocks for a more detailed view of this \citep{Guillet2010,Anderl2013}.\par
The model discussed here has been successfully used to explain the behaviour of molecules like H$_2$O, NH$_3$, CS and H$_2$S in shocks like L1157-B1 \citep{viti2011,Gomez-Ruiz2014,holdship2016}. The parameterization has been extensively tested against MHD C-shock simulations and performs well across a wide range of input parameters. Importantly, it is computationally simple. This allows the chemistry to be the main focus without the computational costs being too prohibitive to run large grids of models to compare quickly to observations.\par
\subsection{Network-MakeRates}
\label{sec:makerates}
A network of all the included species and their reactions must be supplied to the model. For UCLCHEM this consists of two parts: the gas-phase reactions taken from the UMIST database and a user-defined set of grain reactions. Grain reactions include freeze out, non-thermal desorption and any reactions between grain species. The UCLCHEM repository includes the UMIST12 data file \citep{McElroy2013} and a simple set of grain reactions. Given that the code must be explicitly given the ODEs and told which species to sublimate and in which proportions for different thermal desorption events, a python script has been created to produce the input required to run UCLCHEM from a list of species, a UMIST file and a user created grain file. \par

\section{Application I: Collapsing Cores}
\label{sec:collapses}
To provide an example of the capabilities of UCLCHEM, the results of a model run for a collapsing core of gas is presented here. In the earliest stages of star formation, protostars form from collapsing cores of gravitationally bound gas. However, the evolution of the density profiles of these cores is not well understood. UCLCHEM can be used with a variety of theoretical time-dependent density profiles. In a separate paper, Priestley et al. (in prep) will discuss these profiles and their chemical effects in detail. Here, a simple and commonly considered profile is demonstrated, that of a marginally super-critical Bonnor-Ebert (BE) sphere \citep{Bonnor1956}. The link to the code provided in previous sections will contain the most up to date release of UCLCHEM, a permanent record of the code used for this work is available at \url{http://dx.doi.org/10.5281/zenodo.580044}.\par
BE spheres are a solution to the hydrostatic equilibrium equations and have been shown to fit the observed density profiles of prestellar cores. This typically requires that the cores are close to the critical mass which would make them unstable to collapse \citep{Kandori2005}. The cloud model includes a parameterization of simulations by \citet{Foster1993} who consider a BE sphere at the critical mass for stability and then increase the density throughout the sphere by 10\% making it marginally super-critical. Thus, the sphere collapses and the chemistry of a prestellar core collapsing in this fashion can be modelled. Such applications are useful as it may be possible to discriminate between different collapse models from the observed chemical differentiation in the core, even if the density profile cannot be confidently described due to the lack of a sufficiently high angular resolution.\par
\subsection{Model Setup}
The model runs consider 50 gas parcels equally distributed from the core to the edge of a cloud of size 0.2 pc and with an initial density of \SI{2e2}{\per\centi\metre\cubed}. This cloud is embedded in a larger medium which adds 1.5 magnitudes to the visual extinction at the cloud edge. This cloud maintains a temperature of \SI{10}{\kelvin} as it collapses according to the freefall equation,
\begin{equation}
\frac{dn}{dt} = \left(\frac{n_H^4}{n_0}\right)^{\frac{1}{3}} \left[24\pi G m_H n_0 \left(\left(\frac{n_H}{n_0}\right)^{\frac{1}{3}}-1\right)\right]^{\frac{1}{2}}
\end{equation}
where $n_H$ is the hydrogen nuclei number density, $n_0$ is the initial central density and m$_H$ is the mass of a hydrogen nucleus. The initial molecular abundances are set to zero and elements are set to solar abundance values \citep{Asplund2009}. The cloud then collapses until it reaches a density of \SI{2e4}{\per\centi\metre\cubed}, the initial density used for the BE collapse. This takes approximately \SI{3.5}{\mega\year}. The chemical processes described in Section~\ref{sec:chemmod} occur during this collapse. This process is typically referred to as phase 1 in work using UCLCHEM, and is used to create initial abundances and physical conditions for the following phase which is typically the process of interest. In this case, the abundances at the end of this collapse are used as the initial abundances for model runs which start at \SI{2e4}{\per\centi\metre\cubed} and collapse to \SI{1e8}{\per\centi\metre\cubed} in either freefall collapse or according to the BE collapse parameterization.
\subsection{Results}
In order to test the model, it is useful to compare to observations. \citet{Tafalla2004} (T04) present observations of two starless cores, L1517B and L1498, which are well approximated by BE spheres with central densities of approximately \SI{2.2e5}{\per\centi\metre\cubed} and \SI{0.94e5}{\per\centi\metre\cubed} respectively. Though the reported characteristics for each core are very similar, L1498 is used as a comparison to the model here. T04 derive temperature and density profiles from the dust emission. They then derive CS, CO, NH$_3$ and NH$_2^+$ abundance profiles from radiative transfer fitting to the flux profile, favouring the simplest profile that recreated the data.\par
\begin{figure*}
\centering
\plotone{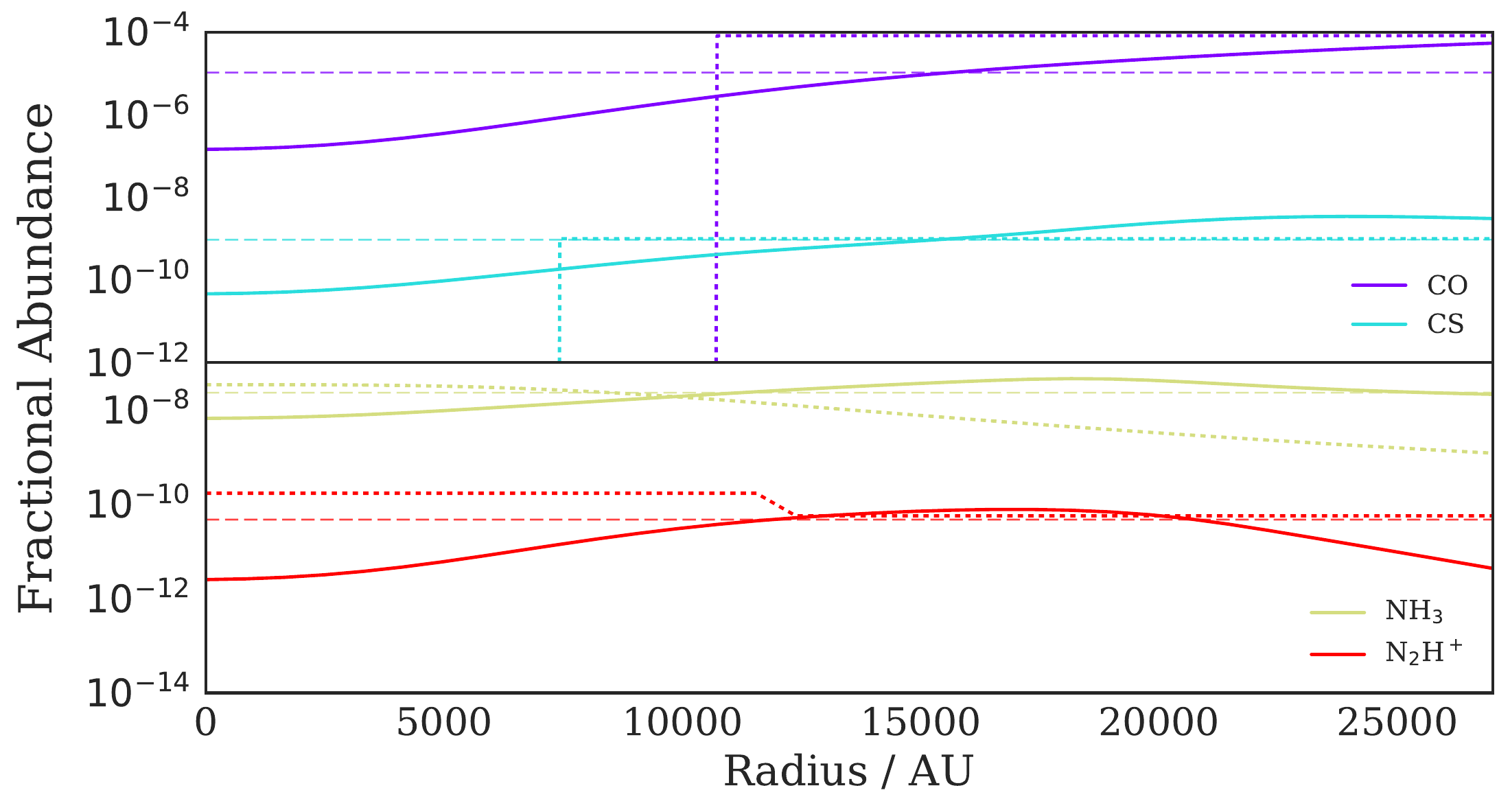}
\caption{Fractional abundance as a function of radius for molecules from \citet{Tafalla2004} modelled by UCLCHEM are plotted as solid lines. Also plotted as dashed, horizontal lines, the results of a freefall model are plotted to demonstrate the improvements to the model that using a BE sphere collapse achieves.\label{fig:beabund}}
\end{figure*}

They find CO and CS to be depleted in the denser centres of the cores compared to the outer radii and that this is adequately represented as a step function. In contrast, NH$_3$ is enhanced by a factor of a few in the centre; a trend they represent as a simple radial power law. N$_2$H$^+$ has a constant abundance for L1517B and a slight drop at large radii for L1498. Chemical models including the one presented here accurately reflect the behaviour of CS and CO. However, many struggle to reproduce the NH$_3$ and N$_2$H$^+$ behaviour \citep{Aikawa2003}. Figure~\ref{fig:beabund} shows results of UCLCHEM using the BE model used to fit these observations with a central density of \SI{0.95e5}{\per\centi\metre\cubed}. \par
From Figure~\ref{fig:beabund} it can be seen that the BE sphere modelling with UCLCHEM meets with some success. For CO and CS, T04 find each can be treated as having constant abundance outside of some cut off radius, with the abundance in the core being greatly depleted. For simplicity, T04 take a central abundance of zero and whilst this is clearly not the case for UCLCHEM, there are some caveats. The T04 observations are of C$^{17}$O and C$^{18}$O and they report C$^{18}$O abundances. For the comparison here, those values have been multiplied by 550, the C$^{16}$O:C$^{18}$O ratio \citep{Wilson1999}. Reducing the CO values by this amount gives a central abundance of 2$\times10^{-10}$, which is sufficiently low that the signal should be too weak to detect. Similarly, the central abundance of CS in the UCLCHEM model is 4$\times10^{-11}$ and therefore in agreement with zero.\par
T04 provide an analytic expression for NH$_3$ which is plotted in Figure~\ref{fig:beabund}, the main finding is that NH$_3$ increases by a factor of a few towards the centre of the core. The UCLCHEM model predicts the opposite, increasing with radius. UCLCHEM similarly struggles to reproduce the observations of N$_2$H$^+$, decreasing both towards the centre and edge of the core rather than the approximately constant abundance found in T04. It appears from the model outputs that at middle radii, N$_2$H$^+$ is forming through a reaction between H$_3^+$ and N$_2$. The increased density of the middle radii makes this reaction more efficient than it is at the edges, but the even higher density and visual extinction of the centre reduces the amount of H$_3^+$ available and further increases the rate of freeze out.  \par

The freefall model is plotted on the same figure using thinner dashed lines. The freefall model only takes into account the initial density of each parcel and not the radius so there is no differentiation with distance from the core centre. Further, the freefall model reaches \SI{1e5}{\per\centi\metre\cubed} much more quickly than the BE model and so the depletion of each species is much lower than expected as freeze out has less time to occur. In many situations, the freefall model is sufficient for single point models that describe the densest part of a cloud of gas but it is clearly inappropriate to model the collapse of prestellar cores. \par
The short comings of the BE model could be due to a number of factors. Spherical symmetry is assumed in both the UCLCHEM model and in the radiative transfer modelling of T04, who fit to azimuthally averaged data. Alternatively, the collapse of a marginally super-critical BE sphere may not be exactly appropriate for this object or the initial conditions for the chemical model may be incorrect. The fact the behaviour of both N$_2$H$^+$ and NH$_3$ are not reproduced by the BE model could indicate a problem with the nitrogen chemistry in the dense gas at core centre. However, the model is clearly superior to a simple freefall collapse and a more in depth investigation into collapse modes and initial conditions could link the observed abundance profiles and the chemical modelling. The strength of UCLCHEM is that it has been specifically designed to have easily modified, and easy to implement, physics modules. To illustrate the code's range of physical applications, we present a second example with the chemical modelling of C-type shocks in the next section.\par

\section{Application II: Shock Chemistry}
\label{sec:shocks}
\subsection{How Can We Trace Shocks?}
In this section, we present a second example use of UCLCHEM that focuses on the study of C-type shocks. A primary concern of chemical modelling is the identification of tracer molecules, the observation of which can provide information on a quantity that is not directly observed. We aim to find molecules that trace shocks, particularly those sensitive to the shock properties. \par%
Here, we build on the work of \citet{viti2011} in which H$_2$O and NH$_3$ lines were used to determine the properties of a shock. Some molecules, such as H$_2$O trace the full shock~as they are not destroyed in the hotter parts of the post shock gas, which then dominates the high-J emission. Others, such as NH$_3$ may not and are instead destroyed in the hotter parts of the post shock gas. This behaviour depends on both the pre-shock density and the shock speed so the determination of which molecules fall into which category in varying conditions is a rich source of information. We provide groupings of molecules under different shock conditions (pre-shock density and shock speed) so comparisons can be made between groups when observing shocked gas.\par
\subsection{Model Setup}
\label{sec:setup}
As explained in previous sections, the model is two phased. Phase 1 starts from purely atomic gas at low density ( n$_H$=10$^2$ cm$^{-3}$ in this work), which collapses according to the freefall equation, in a manner similar to the core collapse model described above. This sets the starting abundances for all molecules in phase 2, which will be self-consistent with the network rather than being assumed values for a molecular cloud. The elemental abundances are inputs and for this work, typical elemental solar abundances from \citet{Asplund2009} were used. The cloud collapses to the initial density of phase 2 and is held there until a chosen time. The grid run for this work used a phase 1 of 6 Myr, which provided sufficient time for the required pre-shock densities to be reached.\par
Table~\ref{tab:parameters} lists the physical parameters that were set for all models ran for this work. The phase 1 models ran here used a temperature of 10 K, a radiation field of 1 Habing and the standard cosmic ray ionization rate of 1.3 $\times$ 10$^{-17}$ s$^{-1}$. A radius of 0.05 pc is chosen for the cloud and this sets the Av by allowing the column density between the edge of the cloud and the gas parcel to be calculated assuming a constant density. An additional 1.5 magnitudes are added to the Av, an assumed value for the amount of extinction from the interstellar radiation field to the edge of the cloud. The final density of this phase varied depending on the density required for phase 2 (See Table~\ref{tab:models}). \par
\begin{figure}
\centering
\plotone{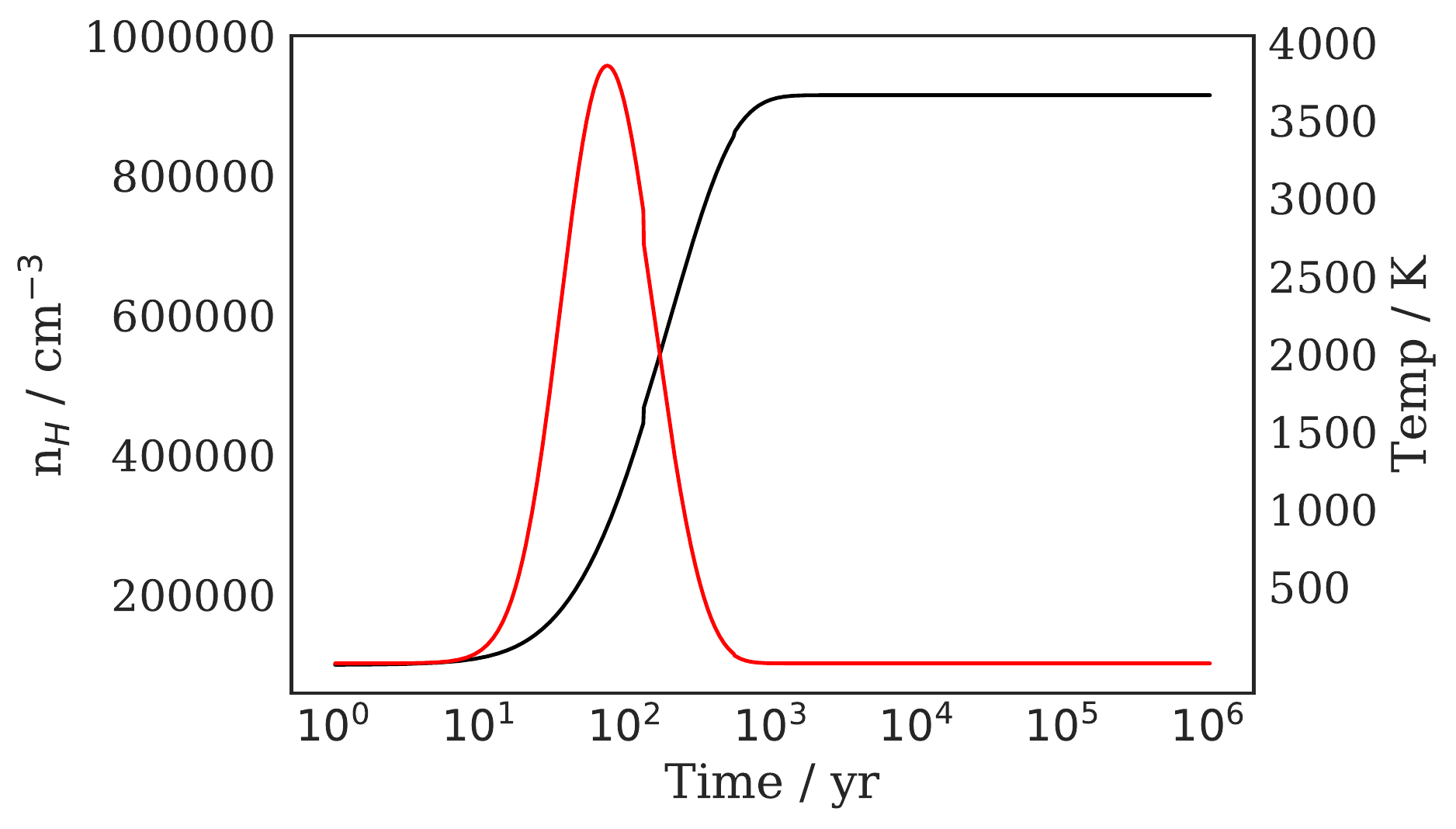}
\caption{Neutral fluid temperature and density profiles of a typical model, in this case model 27 is shown. Temperature is shown in red and the density in black.\label{fig:physcon}}
\end{figure}

In phase 2, the C-shock occurs. Table~\ref{tab:models} gives a full list of the C-shock models that were run. Note, not all of the listed variables are independent, t$_{sat}$ and T$_{max}$ are functions of the pre-shock density (n$_{H,0}$) and shock velocity (V$_s$). The T$_{max}$, V$_s$ relationship is extracted from figures 8 and 9 of \citet{Draine1983} which give values of maximum temperature for pre-shock densities 10$^4$ and 10$^6$ cm$^{-3}$. Since no values are available when the pre-shock density is 10$^5$ cm$^{-3}$ the maximum temperature is calculated as the average of the 10$^4$ and 10$^6$ values. More details of these variables and how they are calculated can be found in \citet{jimenez2008}. The temperature and density profiles of the gas during this C-shock are shown in Fig.~\ref{fig:physcon}, using model 27 as an example.
\begin{table}[]
\centering
\caption{List of Model Parameter Values}
\label{tab:parameters}
\begin{tabular}{ll}
Variable & Value\\
\hline
Pre-shock Temperature & 10 K \\
Initial Density & 10$^2$ cm$^{-3}$\\
Freeze Out Efficiency & 1.0 \\
Radiation Field & 1.0 Habing \\
Cosmic Ray Ionization Rate & 1.3 $\times 10^{-17} s^{-1}$ \\ 
Visual Extinction at Cloud Edge & 1.5 mag\\
Cloud Radius & 0.05 pc \\
Grain Surface Area & 2.4 $\times 10^{-22}$ cm$^2$ \\
$H_2$ Desorption ($\epsilon$) & 0.01\\
CR Desorption ($\phi$) & 10$^5$\\
UV Desorption (Y) & 0.1 \\
\hline
\\
\end{tabular}
\end{table}

\begin{table}[]
\centering
\caption{List of Shock Variable Values}
\label{tab:models}
\begin{tabular}{llllll}
Model & log(n$_{H,0}$) & V$_s$ & t$_{sat}$& T$_{max}$& Z$_{diss}^*$\\
&&/ km s$^{-1}$& / yr & / K & / cm \\
\hline
1	&	10$^3$	&	10	&	975.7	&	323	&	3.70$\times10^{17}$\\
2	&	10$^4$	&	10	&	97.6	&	323	&	3.70$\times10^{16}$\\
3	&	10$^5$	&	10	&	9.8	&	301	&	3.70$\times10^{15}$\\
4	&	10$^6$	&	10	&	1.0	&	279	&	3.70$\times10^{14}$\\
5	&	10$^3$	&	15	&	746.5	&	584	&	5.55$\times10^{17}$\\
6	&	10$^4$	&	15	&	74.7	&	584	&	5.55$\times10^{16}$\\
7	&	10$^5$	&	15	&	7.5	&	555	&	5.55$\times10^{15}$\\
8	&	10$^6$	&	15	&	0.8	&	525	&	5.55$\times10^{14}$\\
9	&	10$^3$	&	20	&	586.13	&	869	&	7.41$\times10^{17}$\\
10	&	10$^4$	&	20	&	58.61	&	869	&	7.41$\times10^{16}$\\
11	&	10$^5$	&	20	&	5.86	&	893	&	7.41$\times10^{15}$\\
12	&	10$^6$	&	20	&	0.59	&	916	&	7.41$\times10^{14}$\\
13	&	10$^3$	&	25	&	483.09	&	1178	&	9.26$\times10^{17}$\\
14	&	10$^4$	&	25	&	48.31	&	1178	&	9.26$\times10^{16}$\\
15	&	10$^5$	&	25	&	4.83	&	1316	&	9.26$\times10^{15}$\\
16	&	10$^6$	&	25	&	0.48	&	1454	&	9.26$\times10^{14}$\\
17	&	10$^3$	&	30	&	425.83	&	1510	&	1.11$\times10^{18}$\\
18	&	10$^4$	&	30	&	42.58	&	1510	&	1.11$\times10^{17}$\\
19	&	10$^5$	&	30	&	4.26	&	1824	&	1.11$\times10^{16}$\\
20	&	10$^6$	&	30	&	0.43	&	2137	&	1.11$\times10^{15}$\\
21	&	10$^3$	&	35	&	402.8	&	1866	&	1.30$\times10^{18}$\\
22	&	10$^4$	&	35	&	40.28	&	1866	&	1.30$\times10^{17}$\\
23	&	10$^5$	&	35	&	4.03	&	2416	&	1.30$\times10^{16}$\\
24	&	10$^6$	&	35	&	0.4	&	2966	&	1.30$\times10^{15}$\\
25	&	10$^3$	&	40	&	402.47	&	2245	&	1.48$\times10^{18}$\\
26	&	10$^4$	&	40	&	40.25	&	2245	&	1.48$\times10^{17}$\\
27	&	10$^5$	&	40	&	4.02	&	3093	&	1.48$\times10^{16}$\\
28	&	10$^6$	&	40	&	0.4	&	3941	&	1.48$\times10^{15}$\\
29	&	10$^3$	&	45	&	413.29	&	2648	&	1.67$\times10^{18}$\\
30	&	10$^4$	&	45	&	41.33	&	2648	&	1.67$\times10^{17}$\\
31	&	10$^5$	&	45	&	4.13	&	3855	&	1.67$\times10^{16}$\\
32	&	10$^3$	&	60	&	397.28	&	3999	&	2.22$\times10^{18}$\\
33	&	10$^4$	&	60	&	39.73	&	3999	&	2.22$\times10^{17}$\\
34	&	10$^3$	&	65	&	337.32	&	4497	&	2.41$\times10^{18}$\\
35	&	10$^4$	&	65	&	33.73	&	4497	&	2.41$\times10^{17}$\\
\hline
\\
\end{tabular}
\begin{flushleft}
* Z$_{diss}$, the dissipation length is the length scale over which the shock dissipates, beyond which the model is no longer applicable.
\end{flushleft}
\end{table}
\subsection{Shock Tracers}
Due to the fact freezing is efficient in cold, dense clouds, many species are highly abundant in the solid phase in the pre-shock cloud model used in this study.  The result is that the fractional abundances of a similarly large number of species increase by orders of magnitude in shocks of any speed and can be used to determine whether a shock has passed. These are typically molecules that do not have efficient formation mechanisms in cold gas but are easily formed on the grain. Hydrogenated molecules in particular tend to fall into this category.\par
\begin{figure*}
\centering
\plotone{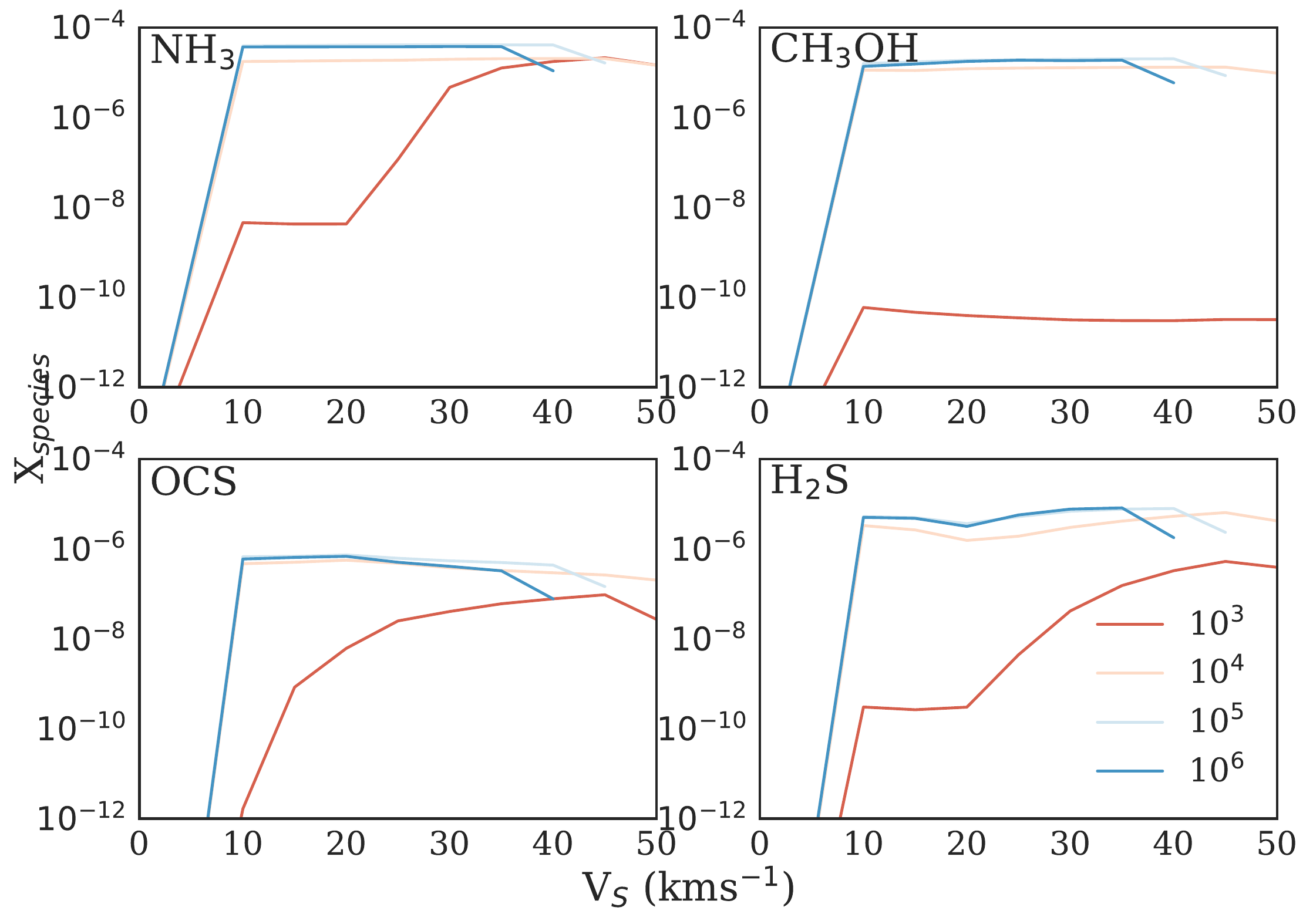}
\caption{Average fractional abundance as a function of shock velocity for a number of molecules that have low gas-phase abundances in the pre-shock gas but show high abundance increases in shocks of any velocity. Each line shows a different pre-shock density, the log(n$_H$) values are given in the lower right plot. In the absence of other mechanisms capable of sublimating most of the species in the ices, high abundances of these molecules would indicate the passage of a shock.\label{fig:simple}}
\end{figure*}

H$_2$O and NH$_3$ are highly abundant in the ices but not the gas phase in cold molecular clouds. Of the two, NH$_3$ is a more useful tracer of shocks due to the difficulties in observing H$_2$O from the ground. In the models, NH$_3$ has a fractional abundance of 10$^{-14}$ in the pre-shock gas but can increase to 10$^{-5}$ in a shock (see Figure~\ref{fig:simple}). H$_2$S and CH$_3$OH similarly go from extremely low abundances in the pre-shock gas to X$ \sim 10^{-5}$ in a shock. Fractional abundances as a function of shock speed for different pre-shock densities are shown in Fig.~\ref{fig:simple}.\par
An important caveat to this is that the model assumes all of the grain surface material is released into the gas phase unchanged. This may not be the case, as the collisions that cause sputtering may have enough energy to destroy molecules as they are released. \citet{Suutarinen2014} discuss this possibility in the context of CH$_3$OH being released from ice mantles. They find that more than 90\% of CH$_3$OH is destroyed either in the sputtering process itself or in gas-phase reactions after sublimation. The latter is the focus of the next section of this work but it is not clear how much each process contributes to the destruction. However, the species presented here increase by over six orders of magnitude in gas-phase abundance due to sputtering, if even 1\% survives the process then the enhancement is observable.\par
An attempt was made to identify species that vary in fractional abundance by orders of magnitude depending on the shock speed. This is in contrast to those shown in Fig~\ref{fig:simple} which show a steep change from no shock to a low velocity shock but no variation thereafter. Such species would allow for the possibility of determining shock properties directly from the abundance of certain molecules. Despite many species increasing in shocks, it was not possible to identify any such species. In the model, even the low velocity shocks (10 km s$^{-1}$) sputter the grains so all shocks can increase gas-phase abundances in this way. Any variability due to gas temperatures reached in the shock is lost when an average over the whole shock is taken, with abundances varying by less than an order of magnitude.\par
It may not be possible to use the abundances directly to determine the shock velocity. However, UCLCHEM provides a useful input for the radiative transfer: a gas-phase abundance for the molecule as a function of the passage of the shock. Further, as discussed in Section~\ref{sec:shockprops}, whilst bulk abundances are not sensitive to shock speed, line profiles can be used to provide further constraints.\par
\subsection{Determining Shock Properties}
\label{sec:shockprops}
The focus so far has been on average abundances across the whole shock, the observed abundance if one just resolves the shocked region. However, molecules can exhibit differences through the shock at different velocities even when the average is unchanged. If these differences are ignored, the models can only be used to infer that a shock has passed. However, if the velocity profile of the emission is taken into account, much more can be learned from the models. \par
For example, \citet{Codella2010} presented NH$_3$ and H$_2$O profiles from the L1157-B1 region and showed that when scaled to be equal at their peak emission velocity, NH$_3$ dropped off much more quickly with increasing velocity. \citet{viti2011} demonstrated that this is consistent with NH$_3$ being at lower abundance in the hotter gas, with chemical models showing that a shock with v$_s \textgreater$ 40 km s$^{-1}$  and pre-shock density of n$_H$=10$^5$ cm$^{-3}$ was sufficient to destroy NH$_3$ which, unlike H$_2$O does not efficiently reform in the gas phase. As such, a divergence in the velocity profiles of these molecules can indicate such a shock.\par
\citet{Gomez-Ruiz2016} took this further and showed the difference in the terminal velocity of NH$_3$ and H$_2$O was correlated with the shock velocity inferred from the H$_2$O terminal velocity for a sample of protostellar outflows. Chemical modelling in the work corroborated this by showing more NH$_3$ being lost in faster shocks. \citet{holdship2016} showed H$_2$S exhibits the same behaviour as NH$_3$, dropping in abundance in high velocity gas. Given the promise of this method, molecules that exhibit this behaviour are explored so this analysis can be easily extended by observing lines of these molecules. It should be noted that the key assumption of the above studies is that the decrease in emission intensity with velocity between two molecules emitting from the same source is due to a similar change in abundance with velocity. Excitation effects will complicate this but when comparing optically thin emission scaled to match at peak velocity, it should hold as verified with radiative transfer modelling in \citet{viti2011}.\par
\begin{figure*}
\centering
\plotone{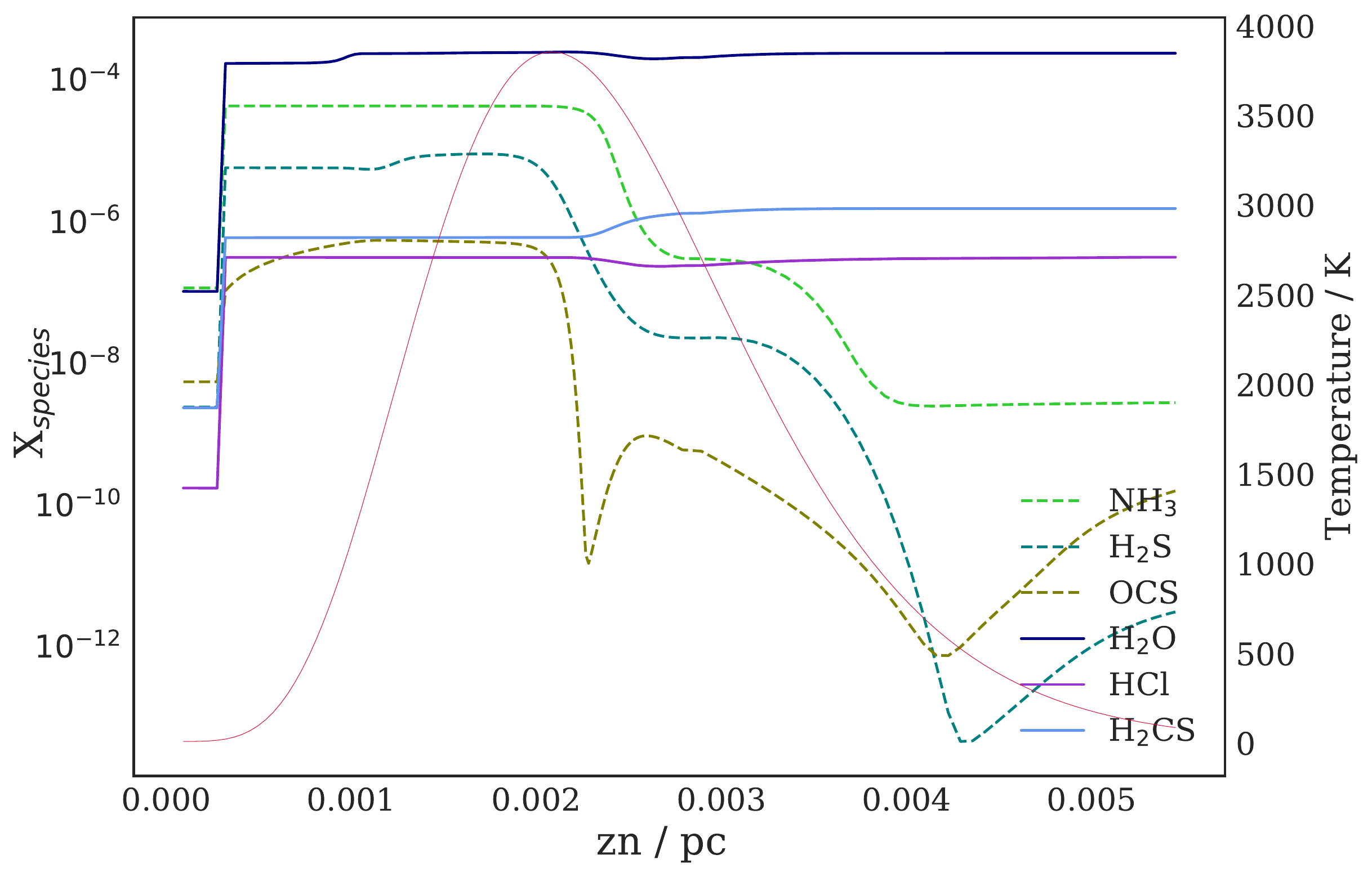}
\caption{Fractional abundances as a function of distance into a shock for model 31. Species from each of the two groups are displayed as an illustration of their differences. The solid lines show H$_2$O-like molecules and the dashed show NH$_3$-like molecules. The neutral gas temperature profile is shown in red to show where the NH$_3$-like molecules start to fall in abundance.\label{fig:model}}
\end{figure*}

NH$_3$ and H$_2$S are unlikely to be the only molecules enhanced in shocks and subsequently destroyed in hot gas. Equally, species other than H$_2$O will remain enhanced throughout a shock and will thus be able to trace the whole post-shock region. H$_2$O can be difficult to observe with ground-based telescopes and so an alternative species showing a similar behaviour could be used as a proxy for H2O. The modelling suggests two groups of molecules: shock-tracing (H$_2$O-like) and shock-destroyed (NH$_3$-like). Fractional abundances through a shock model for selected species from each categories are shown in Figure~\ref{fig:model} to illustrate their behaviour. Observing molecules from each group and comparing their velocity profiles and abundances to the predictions of the model can constrain the shock properties.\par
\begin{figure*}
\centering
\plotone{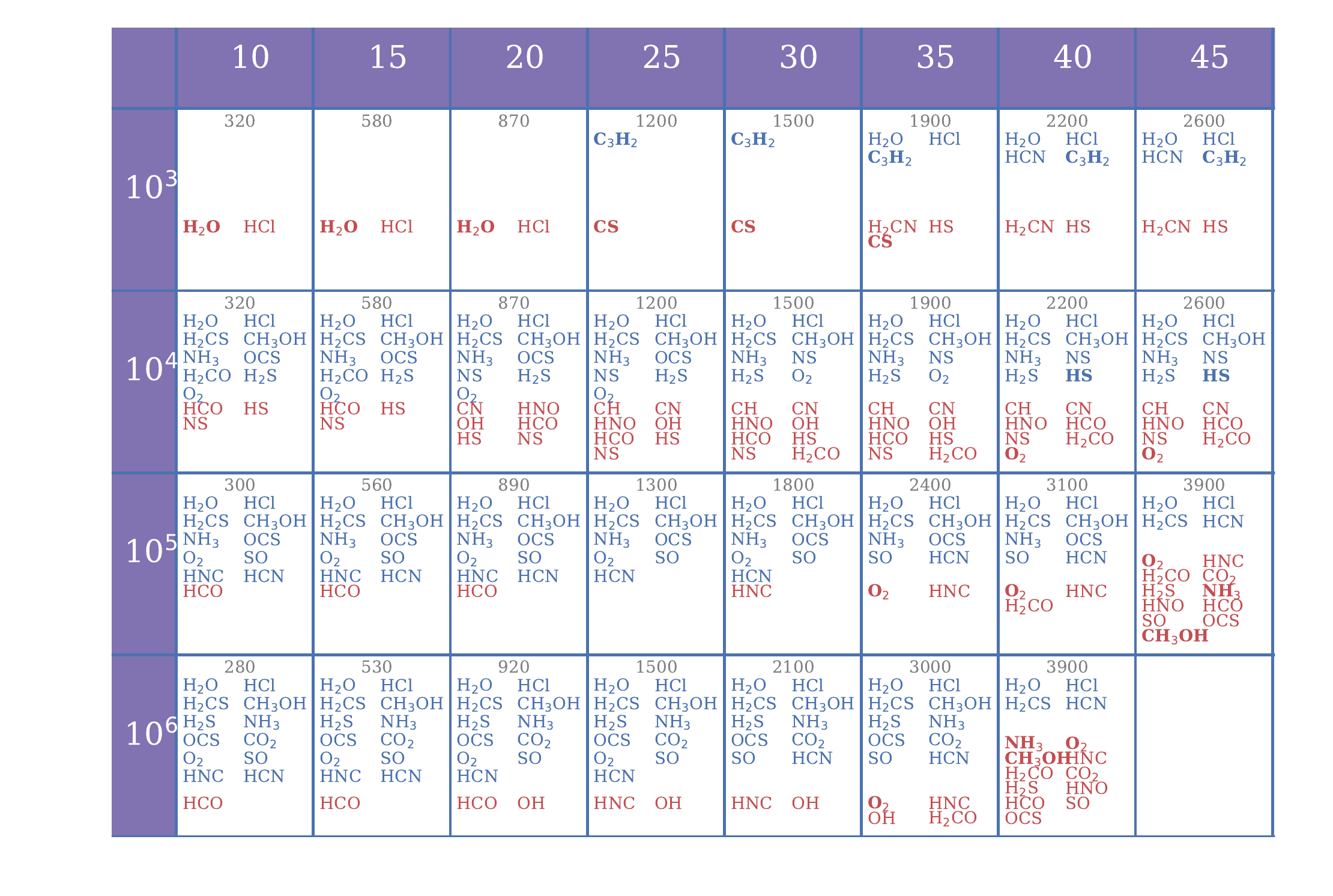}
\caption{Shock tracers for all initial density (vertical values), shock speed (horizontal values) pairs from Table~\ref{tab:models}. The maximum gas temperature reached in each model is displayed in the relevant box. Species in blue are enhanced by the shock and trace its full extent. Species in red are enhanced initially by the shock but are destroyed as the shock increases the temperature. Species in bold are those which exhibit a given behaviour only for a small range of conditions, making them particularly useful for determining shock properties. The bottom right panel is left blank, a C-type shock cannot propagate at 45 km s$^{-1}$ through a medium of density 10$^6$ cm$^{-3}$.
\label{fig:tracers}}
\end{figure*}

A species does not necessarily fit one category or the other for all shock conditions. The behaviour is the result of temperature and density profiles of the shock and so depends on the shock speed and initial density. In Figure~\ref{fig:tracers} all modelled pairs of shock speed and initial density are shown with lists of shock-tracing and shock-destroyed species in blue and red respectively. Species which exhibit more complicated behaviour and cannot be easily categorized are simply omitted. Some species, such as H$_2$O are almost always found to fall in one category, this allows them to be consistently used for a given purpose. ie. H$_2$O can be used to trace the full shock in most cases. However, some species will demonstrate a certain behaviour only in very specific conditions and can be used to limit possible shock properties when this is observed. For example, H$_2$O does not trace the full shock if the initial density is 10$^3$ cm$^{-3}$ and the shock is low speed. Such species are displayed in bold in Figure~\ref{fig:tracers}.  \par
The shock-destroyed and shock-tracing molecules are both enhanced by the sputtering caused by the shock. This is important as molecules that show low abundances in the pre-shock gas are unlikely to have large contributions from unshocked gas when a shocked region is observed. For example, CO shows perfect shock-tracing behaviour but is also abundant in the pre-shock gas and its abundance is rather insensitive to shock interaction. As chemistry is the focus for this work, such species are omitted. However, note that high-J transitions of species such as CO are unlikely to be excited by ambient gas and so can be reliably used to trace the full extent of the shock.
\section{Summary}
\label{sec:summary}
A publicly available, time dependent gas-grain chemical model, UCLCHEM, has been described in detail. The code solves the coupled system of ODEs that represents the chemical network when using the rate equation method of modelling chemistry. Various non-thermal and thermal desorption mechanisms are implemented along with gas phase reactions and a simple grain network. The physical models that can be used to control the gas conditions that in turn affect the chemistry are also described. \par
We presented the code via two applications of the new UCLCHEM: a simple prestellar collapse model which represents a typical use of the code, and an investigation into molecular tracers for C-type shocks. This code is available at https://uclchem.github.io and is actively maintained and developed by research teams at UCL and Queen Mary's College London. Diffusion chemistry on the grain surfaces and improvements to the shock model implementation are in progress and users will benefit from continued development.\par

\section{Acknowledgements}
J.H. is funded by an STFC studentship. I.J.-S. acknowledges the financial support received from the STFC through an Ernest Rutherford Fellowship (proposal number ST/L004801/2). The authors would like to thank D. Quenard and A. Coutens for helpful discussions that contributed to the code.
\bibliography{uclchem} 
\end{document}